# Real Time Position Location & Tracking (PL&T) using Prediction Filter and Integrated Zone Finding in OFDM Channel

SHAKHAKARMI NIRAJ
DHADESUGOOR R. VAMAN
Department of Electrical and Computer Engineering
Prairie View A & M University (Member of Texas A & M University System)
Prairie View, Houston, Texas-77446, USA
e-mail: niraj7sk@yahoo.com, drvaman@pvamu.edu, http://cebcom.pvamu.edu

*Abstract:* - The nature of pre-determined and on-demand mobile network fabrics can be exploited for real time Position Location and Tracking (PL&T) of radios and sensors (nodes) for Global Positioning System (GPS) denied or GPS-free systems. This issue is addressed by a novel system of integrated zone finding and triangulation method for determining the PL&T of nodes when mobile network fabrics are employed based on using directional antennas for radio communications. Each mobile node is switched dynamically between being a reference and a target node in PL&T operation to improve the PL&T accuracy of a target node. This paper presents the Baseline PL&T with predictive Kalman filter and Integrated Zone based PL&T algorithm design that integrates zone finding and triangulation method. The performance of the proposed algorithm is analysed using Interleaving-KV sample coding & error correction in Rayleigh and Rician channel using Orthogonal Frequency Division Multiplexing (OFDM) system under the severe multipath fading.

## 1. Introduction

Mobile ad hoc network architectures can be flexibly deployed and the nodes are highly mobile to facilitate supporting a wide variety of emergency disaster scenarios. In some instances these nodes can be air dropped and configured into a set of clusters and allow immediate network operation to support multi-service data exchange. However, these network architectures tend to be bandwidth and resource constrained. They need to be managed skill fully so as to minimize the power of processing and overhead transmission to extend the life time of nodes and allow maximum bandwidth usage for supporting end user applications. Also, while preserving the life time of the nodes, it is important to consider minimization of transmission power to support a target data rate between peer-to-peer nodes. The use of directional antenna enables the power to be focused over the particular zone to provide longer range compared to that of using Omni-directional antenna [1-3], [6]. We exploit the focused coverage of directional antenna to allow detection of the zone in which the target is residing prior to using triangulation for deriving the Position, Location and Tracking (PL&T) of the target node. Universally, tracking of any device in networks has been well established by the use of Global Positioning System (GPS). However, the use of GPS is not secured and in some instances, the use of GPS can be denied. Also, GPS cannot work accurately indoors or near to the buildings and it cannot detect the devices (also referred to as radios) in multi-floor buildings. Therefore, the tracking algorithm needs to rely on the use of reference radios to find PL&T of a target radio(s). Three reference devices are required to track a target radio in (X, Y) plane using triangulation. Similarly, four reference devices are required for tracking a target radio in (X, Y, Z). Since the radios move, the accuracy of tracking suffers due to multiple moving target locations of the radio. Thus, the triangulation method uses re-initialization of the radio location with a known location by moving the target to that location to improve the accuracy. Also, if the reference radios move in the network, the reference radios also suffer from accuracy of their locations. Thus, in a real network centric operation, radios are needed to be switched dynamically to act as

N. Shakhakarmi and D.R.Vaman are with the ARO Center for Battlefield Communications (CebCom), ECE Department, Prairie View A&M University (Texas A & M University system), Prairie View, Houston, Texas-77446, USA.

reference radios or target radios at different instances of time. In addition, each disaster scenario is different from the other and it is important to establish whether fixed reference radios are used or moving reference radios are used. Also, the feasibility of using only external reference radios for tracking indoors or a combination of outdoor-indoor reference radios for tracking indoors. Based upon the decision, it is possible to determine whether Dynamic Switching of Radios (DSR) between being a target radio and being a reference radio for deployment [1-2]. DSR model requires a little bit more processing power and overhead time for the management function.

This paper describes a novel Position, Location and Tracking (PL&T) algorithm based on Time of Departure (ToD) and Time of Arrival (ToA) measurements for each Internet Protocol (IP) packet exchanged between a reference radio and a target radio for determining the range between a reference radio and a target radio. Also using multiple references with known PL&T, the range is translated to [X, Y] co-ordinates for 2D tracking and [X, Y, Z] co-ordinates for 3D tracking using triangulation. It uses KV Transform Coding which is based on orthogonal transformation of four discrete samples into four coefficient samples for transmission. Each discrete sample is creating using n-bits of input data and when transformed, it produces each coefficient samples that can be transmitted with 4 bits using any digital modulation technique. An ensemble of blocks (referred to as KV blocks) with each having four discrete coefficient samples each carrying n-bits are created for transmission. The ensemble of coefficient samples of M-KV blocks is transmitted to the receiving side, where each KV block corrects for 1 out of 4 discrete samples. The remaining KV blocks that have errors due to channel noise are retransmitted exactly once selectively in the next ensemble since the receiver has a knowledge of the location of the KV blocks in error within the ensemble received. In addition, each set of discrete samples are interleaved to reduce the impact of burst errors. It has been shown that for $E_b/N_o$ of << 10 dB, we can recover data at a BER of $10^{-7}$[8] in a multi-path faded channel. The performance of the proposed integrated zone finding and triangulation method is presented while minimizing the impact of multi-path fading and other interference.

We have already researched on PL&T deploying zone forming and triangulation using two reference nodes [1-2]. In this paper, the focus is a single reference node based PL&T and comparison with prediction filter based PL&T for mobile radio nodes. Neighbouring radios are used as references to track a target radio by using the directional coverage of the directional antenna’s beam and the angle of arrival of the directional beam to compute the distance (or range) of the target radio from the reference unit and the angle of the beam between the reference and the target with respect to the baseline of a pair of reference radios [3]. Once the PL&T location of the target is determined, a Kalman filter is employed recursively to predict the next PL&T locations based on error covariance that computes Kalman gain and determine the corrected true position of the radio and the covariance error [4-7]. These true positions are translated into simultaneous localization and map building based on constrained state estimation algorithm [9]. The recursive prediction is continued until the target radio goes out of tracking range.

When GPS is available at each radio, the tracking of the target radio is simpler as it provides GPS data after silent mode. For this case, once the transmitter finds the target, it forms a directional beam and transmits a Request to Send (RTS) message to the target. The target sends a Clear to Send (CTS) message as well as GPS data to the transmitter after it completed the data transfer and goes into silent mode. Target receives all of the data from the transmitter and transmitter performs the Location Prediction Algorithm using Directional Communication (LPADC) only in the forward direction [6]. The limitation of this approach is the unknown silent mode duration and the ability to stay within the coverage area to get the necessary to send GPS data. To address this limitation, researchers have allowed the system not to send CTS until the receiver is in the coverage area and has the ability to send GPS data. Another limitation is that the position prediction is done considering only the straight forward movement and does not consider any sharp turns or obstructions. This is addressed by developing a possible tracking region, formed using the joint information of possible forward movement and sharp turns of the target, based on two previous positions. It is updated with the latest GPS data until the target reaches the coverage boundary. This system can be employed outdoors where reasonable accuracy of GPS data is available, but this cannot be applied both indoors and indoor-outdoor moving radio applications due to severe multi-path interference that effectively minimizes the communications availability. Emergency disaster management applications require radios to move both indoors and outdoors. Even in outdoor where large buildings exist, GPS data many not be very accurate, thus limiting its usage.

When GPS is not accurate, the reliance to PL&T triangulation method using neighbouring references to track target radios is high. Many researchers have demonstrated the use of directional antennas for increasing the coverage area and use signal strengths and the arriving angle of the signal with an established base of a pair of references whose locations are known in an adhoc network environment [1-3]. To improve the accuracy, Kalman filter can be employed in this method, similar to the one used in GPS based algorithm to improve the accuracy of PL&T measurement [4],[7],[9].This algorithm limits its use for motion of the radio with limited directional change. A Minimal Contour Tracking Algorithm (MCTA) is employed to concentrate tracking area where the target vehicle most frequently appears [5]. Although MCTA saves power consumption from the sensor communication, the mapping process for tracking the target with sensors is computationally complex and does not work in high speed vehicle. The energy contour formed by target vehicle can be interrupted or overlapped immediately by other vehicles, which contradicts the MCTA.

## 2. Problem & Proposed Solution

The problem is to formulate robust PL&T scheme using predictive Kalman filter and zone finding approach with triangulation. These PL&T schemes need to be compared under different circumstances. The bit error performance of zone finding method needs to be optimized using OFDM system under severe multipath fading for indoor environment.

The proposed solution includes the following aspects:

- Specifying and executing a Baseline method based PL&T using predictive Kalman filter and triangulation
- Specifying and executing a single reference node based novel PL&T using zone finding and triangulation
- Comparison of different PL&T schemes
- Performance analysis of novel PL&T using interleaving KV transform coding in OFDM based system under multipath fading

### 2.1 Baseline PL&T using Directional Beam with Predictive Kalman Filter

The PL&T requires cross-layer management to support between IP Layer and Physical Medium Dependent (PMD) Layer as shown in Fig.1. At the IP layer, IP Packets are generated for exchange between two radios. For PL&T operation, the IP packets are generated by the PL&T control function in the management when the target is not in communication with the reference radio or vice-versa. When the reference radio and the target are exchanging IP packets as part of normal data transfer, they are used for PL&T operation. The PL&T control sets the number of IP packets in an ensemble and they are time stamped at the PS sub-layer. The Time of Departure (ToD) of each IP packet in the ensemble is recorded. At the remote radio, the Time of Arrival (ToA) of each IP packet in the ensemble is recorded. Because packets get random delays, any packet that arrives after the completion of the ensemble time is not account for ToA measurement. The ToAs of packets received within the ensemble time are transmitted in a management packet to the sender. The sender computes the range based on the difference between ToA and ToD for each packet and the average of all the differences provide the range in particular direction which is the distance as indexed to propagation time. Specifically, directional antenna mode is used by transmitter and omni directional antenna as well as directional antenna mode by receiver to compute ToA, ToD and AoA. This strategy is more efficient in realtime battlefield, when the directional change of node is in the range of -30 to +30 degree until there are twists and turns. Additionally, it provides higher security as the prediction and tracking is based on a single hop only in particular directional range.

In this method, using the directional antenna with a beam width of theta degrees is used to find the range from two references to a target radio using triangulation. The directional antenna is moved in each reference unit, until the target radio is in the beam width of each reference unit and can communi-cate with the two references. Then the range and the corresponding [X,Y] co-ordinate of the target radio is computed to identify the initial location. An extended Kalman Filter is used for recursive prediction, computation and correction of the future

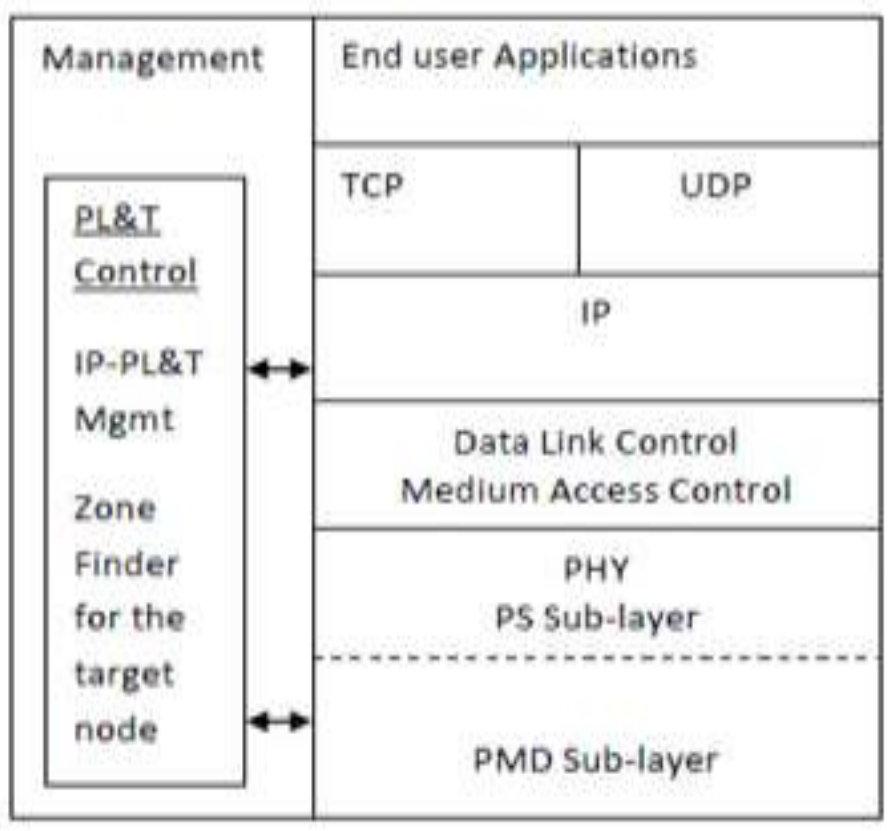

Fig. 1 Radio Architecture for PL&T Computation

PL&T of the target radio over time based on the limited directional path of the target radio and its availability in both beams of the references. Once the target radio is outside one of the beam, then a decision is taken that it is out of range and two new reference nodes are recruited.

The co-ordinates of reference node $R(x_i,y_i)$ which is taken as (0,0) and co-ordinates of transmitting node $S(x_j,y_j)$ which is taken as (x, y) depending upon the received signal strength as shown in Fig. 2. Then, Angle of Arrival (AoA) of signal from R(α) and AoA of signal from S(β) are computed. By triangulation method, coordinates of the desired node $P(x_k,y_k)$ which is being tracked can be computed from the general equation of line PR and PS for the initialization as follows and can be used as reference point for other unknown node in multi-hop scenario as shown in Fig. 2 and equations (1) and (2).

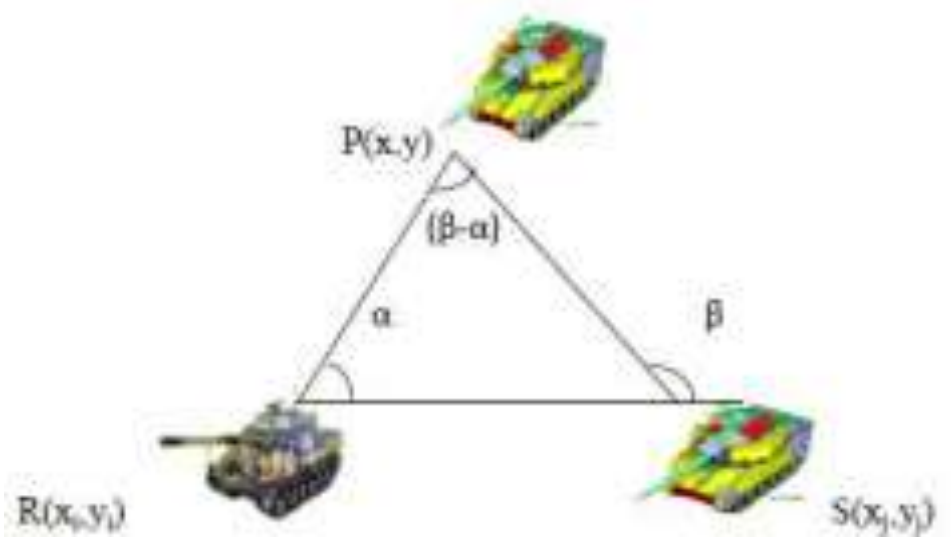

Fig. 2 Illustration of co-ordinate computation

$$x_k=((y_j-y_i)+x_i\mathrm{Tan}\alpha-x_j\mathrm{Tan}\beta)/(\mathrm{Tan}\alpha-\mathrm{Tan}\beta) \quad (1)$$

$$y_k=((x_i-x_j)\,\mathrm{Tan}\alpha\mathrm{Tan}\beta+(y_j\mathrm{Tan}\alpha-y_i\mathrm{Tan}\beta))/(\mathrm{Tan}\alpha-\mathrm{Tan}\beta) \quad (2)$$

Once the initial location is achieved, the future position is derived by the Kalman filter based tracking method which recursively performs the position prediction, estimation and tracking. The discrete Kalman Filter can be modelled as extended Kalman filter since the dynamic equations are linear but the measurement equations are nonlinear to estimate the state vector [4],[7],[9].

i) Prediction of state vector, measurement observation vector and error covariance are computed as follows:

$$X_{k+1}=F_kX_k+B_kU_k \quad (3)$$

$$Y_{k+1}=H_{k+1}X_{k+1}+R_{k+1} \quad (4)$$

$$P_{k+1/k}=F_kP_{k/k}F_k^{\,T}+Q_k \quad (5)$$

ii) Computation of Kalman Filter Gain,

$$S_{k+1}=H_{k+1}\,P_{k+1/k}\,H^T_{k+1}+R_{k+1} \quad (6)$$

$$G_{k+1}=P_{k+1/k}\,H^T_{k+1}\,S^{-1}_{k+1} \quad (7)$$

iii) Corrected error covariance, true state position and recursively continue to (i), (ii) and (iii).

$$P_{k+1}=P_{k+1/k}-G_{k+1}H_{k+1}P_{k+1/k} \quad (8)$$

For unconstrained Kalman Filter;

$$X_{k+1/k}=X_{k+1}+G_{k+1}*(Y_{k+1}-H_{k+1}*X_{k+1}) \quad (9)$$

For constrained Kalman Filter;

$$X_{k+1/k+1}=X_{k+1/k}-P_{k+1/k}*D^T_k*\mathrm{Inv}(D_k*P_{k+1/k}*D^T_k)*D_k*X_{k+1/k} \quad (10)$$

Hence, $X_k$ represents the initial state, $F_k$ is the state transition model, T is the length of the tracking update time interval, $G_k$ is Kalman gain and $P_k$ refers to the estimation error covariance. Similarly, $Y_k$ is the measurement observation, $U_k$ is the control input, $H_k$ is the measurement matrix, $Q_k$ is the process noise covariance, $R_k$ is the measurement noise covariance, $D_k$ is the state constraint matrix, $B_k$ is the input matrix and θ is the heading angle. The receiving node's vehicle dynamics and measurements can be initialized as follows:

$$B_k=\begin{bmatrix}0\\0\\T*\mathrm{Sin}\theta\\T*\mathrm{Cos}\theta\end{bmatrix},\ F_k=\begin{bmatrix}1&0&T&0\\0&1&0&T\\0&0&1&0\\0&0&0&1\end{bmatrix},$$

$$H_{k+1}=\begin{bmatrix}1&0&0&0\\0&1&0&0\end{bmatrix},$$

$$D_k=\begin{bmatrix}1&-\mathrm{Tan}\theta&0&0\\0&0&1&-\mathrm{Tan}\theta\end{bmatrix}$$

Thus, Kalman filter dynamically predict the covariance in estimated position in certain direction with some speed, compute the Kalman gain depening upon actual measurements and provides the corrected true state postion and covariance which is used to predict position in recursive order. In the case of constrained Kalman filter, state constrained is deployed so that position estimate errror can be reduced as compared to the state unconstrained filter case. In some circumstances, once the reference node goes out of radio range from the transmitter then the next neighbouring node is recruited as the reference node and the algorithm needs to work from the beginning.

### 2.2 Integrated Zone based PL&T

The integrated zone is formed considering the intersection of zone for forward movement and zone for sharp turns. This method has used only one reference where as the modified PL&T has used two reference nodes as well as mapping the received energy into radii distance in the reference paper [1]. The position prediction and tracking of desired receiver node can be done by predicting the tracking zone which is created with some geometrical applications as illustrated in Fig. 3. The tracking zone is formed considering the last two positions of the desired node and includes the joint information of forward movement and sharp turns for obstructions. This tactics also needs directional antennas and omni directional antenna for each node and saves drastic amount of power. Directional Antenna mode is used by transmitter and directional antenna as well as directional antenna mode by receiver [6]. This algorithm is most efficient as it provides the tracking zone even the directional change of node is made more than range of -30 to +30 degree for twists and turns on obstructions in the battlefield. It is also more precise and secured as the prediction and tracking based on single hop achieved from the computation of arrival time. The algorithm works as follows:

i) When the last two locations $A(x_{i-1},y_{i-1})$ and B $(x_i,y_i)$ of a receiving node is achieved through CTS signal with GPS data then a line is sketched passing through known A and B. The formulation of this line is $y = mx+ c$, where m is the slope and c is the y-intercept are calculated by following equation:

$$m = (y_i - y_{i-1}) \ / \ (x_i - x_{i-1}) \quad (11)$$

$$c = (x_i y_{i-1} - x_{i-1} y_i) \ / (x_i - x_{i-1}) \quad (12)$$

ii) An equilateral triangle can be formed such that one vertex is at B position and its center $(x_{i+1}, y_{i+1})$ is on line l using following formula and the length of edges of this equilateral triangle is $d_i$ which is the distance between last two locations A and B.

$$x_{i+1} = x_i + \{ d_i * (x_i - x_{i-1}) * \text{Cos}(\arctan(|k|)) \ / |x_i - x_{i-1}| \ \} \quad (13)$$

$$y_{i+1} = y_i + \{ d_i * (y_i - y_{i-1}) * \text{Sin}(\arctan(|k|)) \ / \ 2|y_i - y_{i-1}| \} \quad (14)$$

iii) A circle is drawn that can include the equilateral triangle with smallest radius such that the circle and the equilateral triangle have the same center $(x_{i+1},y_{i+1})$ and the radius of the circle is $r_i$;

$$r_i = d_i / \sqrt{3} \quad (15)$$

iv) Again, a new circle is drawn with radius $r_j = d_i/2$ at the point B $(x_i,y_i)$ and the point P(x,y) where the line joining the centers of two circles meets the line containing the points of intersection of the two circles is computed as follows:

Equation of first circle with radius $r_i$ and centre $(x_{i+1},y_{i+1})$ is given by:

$$(x - x_{i+1})^2 + (y - y_{i+1})^2 = r_i^2 \quad (16)$$

Equation of second circle with radius $r_j$ and centre $(x_i,y_i)$ is given by:

$$(x - x_i)^2 + (y - y_i)^2 = r_j^2 \quad (17)$$

The point P(x,y) where the line joining the centers of two circles meets the line containing the points of intersection of the two circles is computed.

$$x = 0.5*(x_{i+1} + x_i) + \left[ (x_{i+1} - x_i)(r^2_i - r^2_j) / \ 0.5*\{(x_{i+1} - x_i)^2 + (y_{i+1} - y_i)^2\} \right] \quad (18)$$

$$y = 0.5*(y_{i+1} + y_i) + \left[ (y_{i+1} - y_i)(r^2_i - r^2_j) / \ 0.5*\{(x_{i+1} - x_i)^2 + (y_{i+1} - y_i)^2\} \right] \quad (19)$$

Then, draw a circle at point P (x,y) with radius $r_i$ where the future location of neighbor node is predicted and beam width must be defined.

v) Compute the needed beam width $\alpha = 2 \ \text{Arcsin}(r_i / D_j)$ where, $r_i$ is radius of circle and $D_j$ is the distance from transmitter to P(x,y) the point of intersected lines. The real-time beamwidth deployed in Figure-2 is given as:

$$\text{Beamwidth} = \Theta_1 \text{if } \alpha <= \Theta_{1;}$$

$$\Theta_i \text{if } \Theta_{i-1} <= \alpha <= \Theta_{i;}$$

$$\Theta_n \text{if } \alpha >= \Theta_n$$

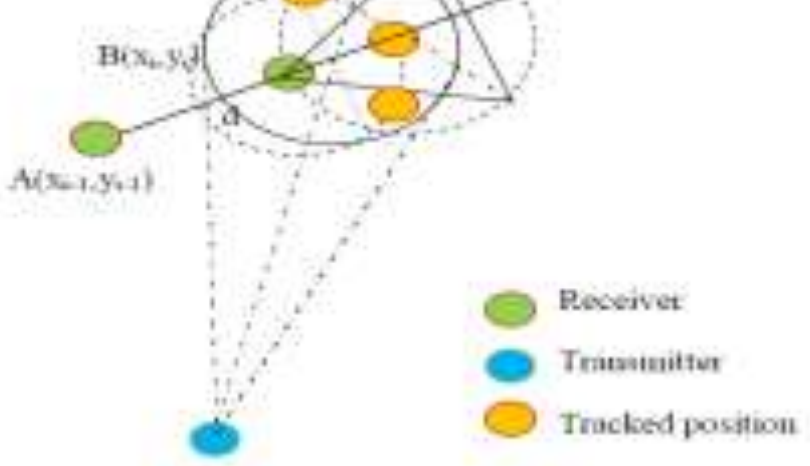

Fig. 3 Adaptive Beam forming over integrated zone

Once the desired node moves close to the boundary of tracking zone then the algorithm must be repeated depending upon last two positions achieved from GPS data.

## 3. Simulation and Performance Evaluation

The simulation is done in C and Matlab for Baseline PL&T and Integrated Zone based PL&T for GPS free systems respectively in 500 X 500 sq. m area. For Baseline PL&T, a transmitter node first selects the neighbouring node as reference node and locates the receiver node which is to be tracked by simply triangle formation with the ToA and ToD computation and angular information. Then, receiver node's future position is predicted in its northern and eastern positions as the battlefield mobility in forward direction is +30 degree to 30 degree. The dynamic equations are linear but the measurement equations are nonlinear, so the extended Kalman filter is deployed to estimate the state vector. Inside the beam coverage within certain range after locating receiver node then it is to be tracked and the covariance of the process and measurement noise are set as:

$$Q(k)= Diag[4m/s, 4m/s, 1m/s^2, 1m/s^2],$$
$$R(k)= Diag[900\ m^2, 900m^2]$$

When the receiver node vehicle is travelling off-road, or on an unknown road, terrains then the state position prediction is complex and this problem is unconstrained. Rest of the case, it may be known that the vehicle is travelling on a given road, which is known as the constrained state estimation using G=Inverse (P). When the vehicle is travelling on a road with a heading angle 60 degree, sample period T is 3 sec and preferred acceleration is set to 1 $m/s^2$. The initial conditions for state vector and error variance are set as expressed in X(k)and P(k/k).

$$X(k)=[0\ 0\ 17\ 10]^T, \qquad P(k/k)=Diag[900\ 900\ 4\ 4]^T$$

The true position profile of receiver node is tracked in North and East direction as shown in Figure-3 and position estimation errors are calculated and simulated in both directional positions using constrained and unconstrained Kalman filter as in Fig. 4-8. The simulation results show that the position estimation error is found lower in constrained filter rather than unconstrained because the initial state is in always favoured as moving along the predefined road in the constrained filter.

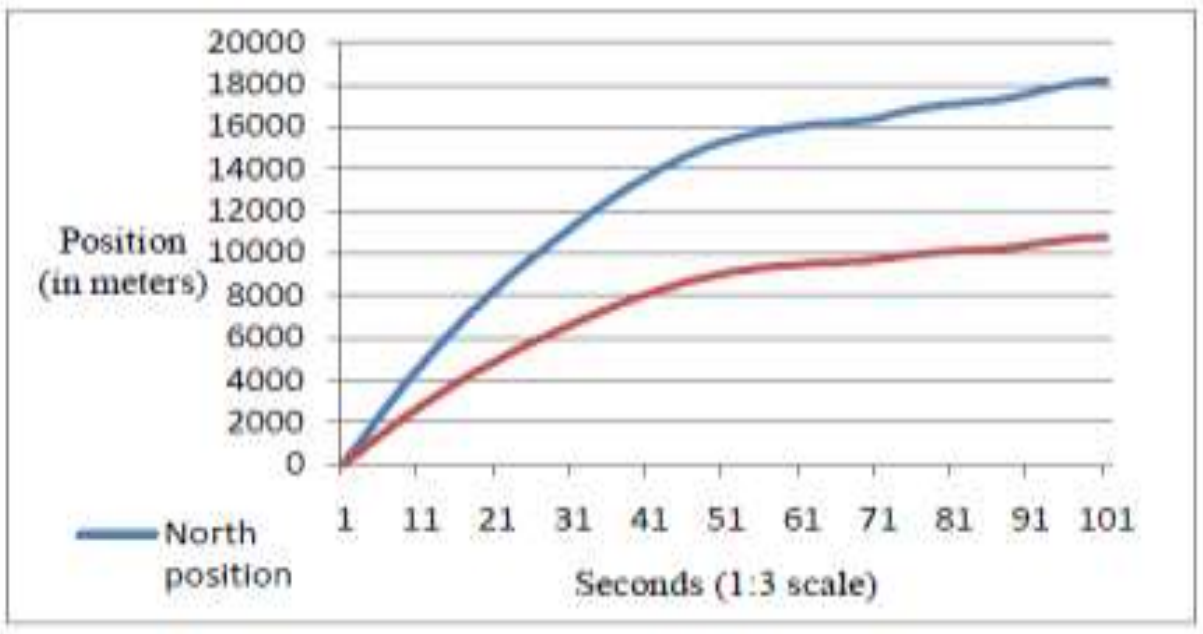

Fig. 4 True position in the North & East directional position.

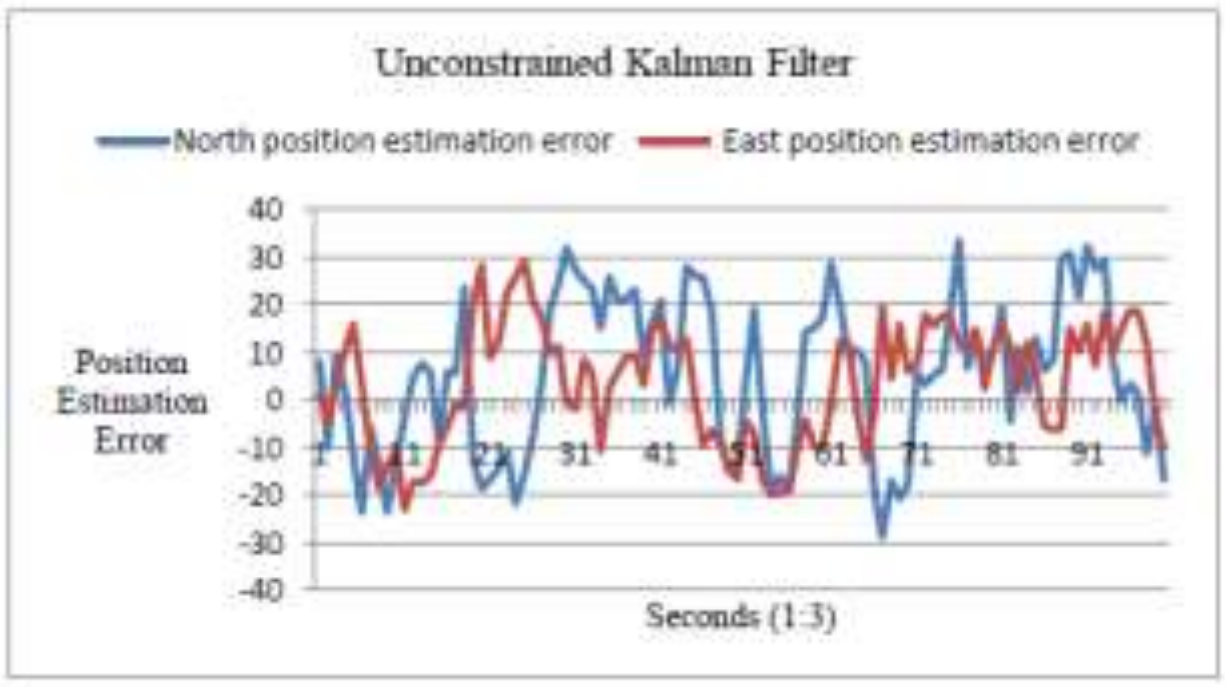

Fig. 5 Unconstrained Kalman Filter position error in the North position and East position.

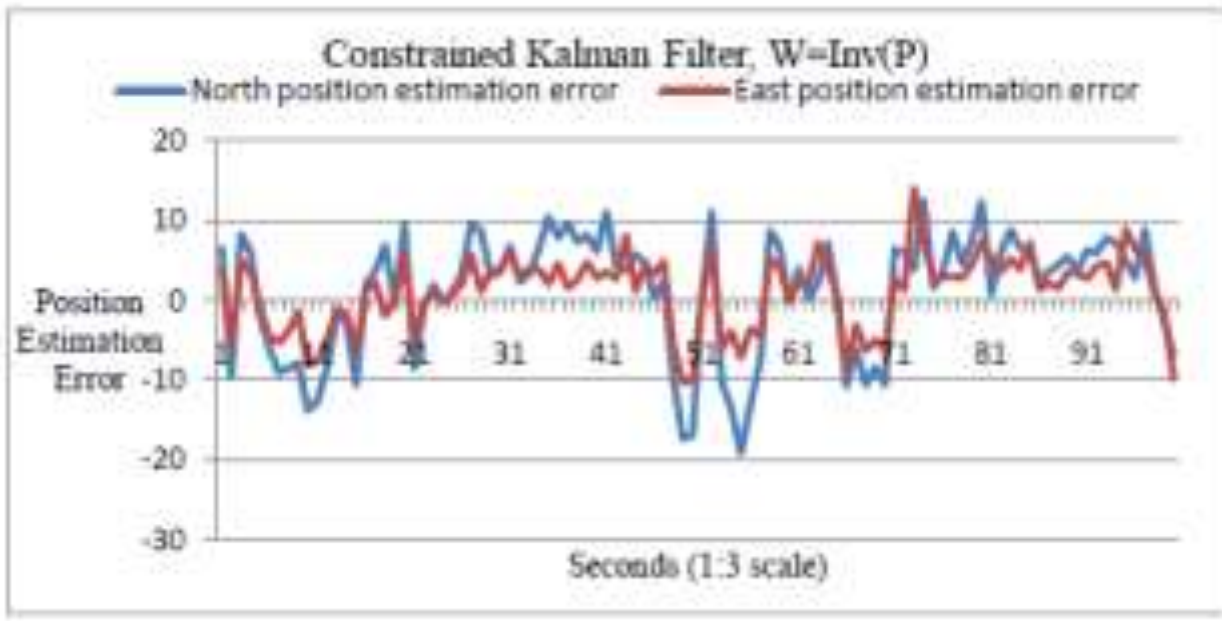

Fig. 6 Constrained Kalman Fillter position error in the North position and East position

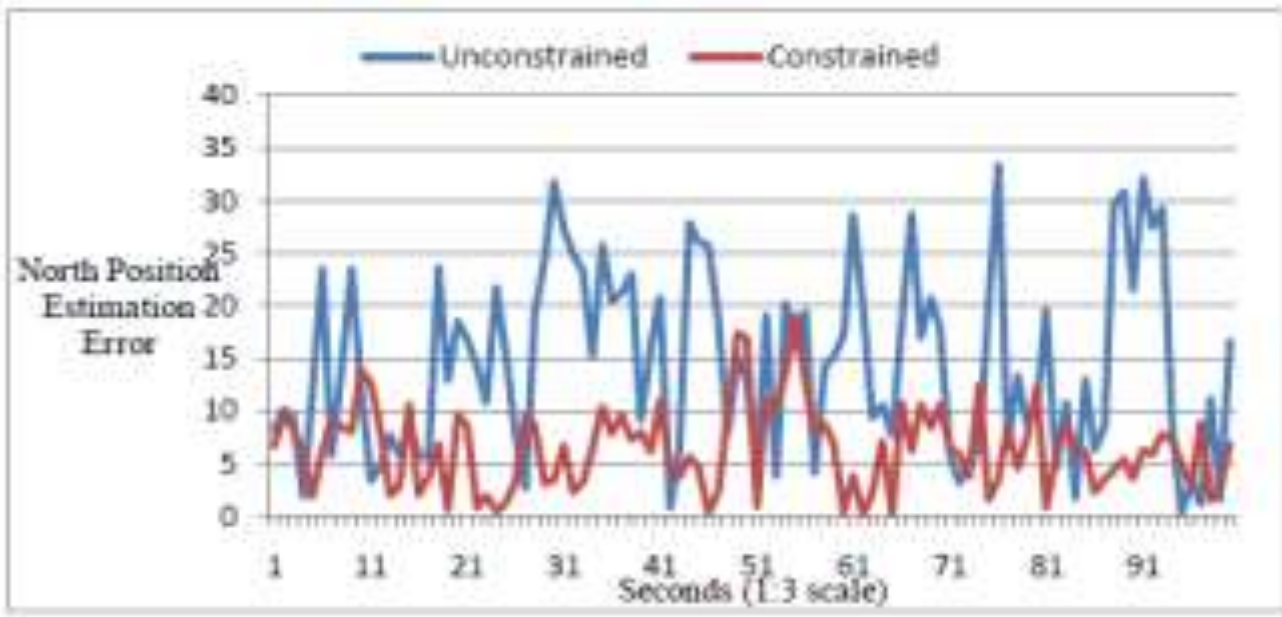

Fig. 7 Unconstrained and constrained KF in the North position estimation error

The average position error is about 5 m in the North position and 3m in East position for unconstrained filter whereas about 1m in both the North and East positions for constrained filter.

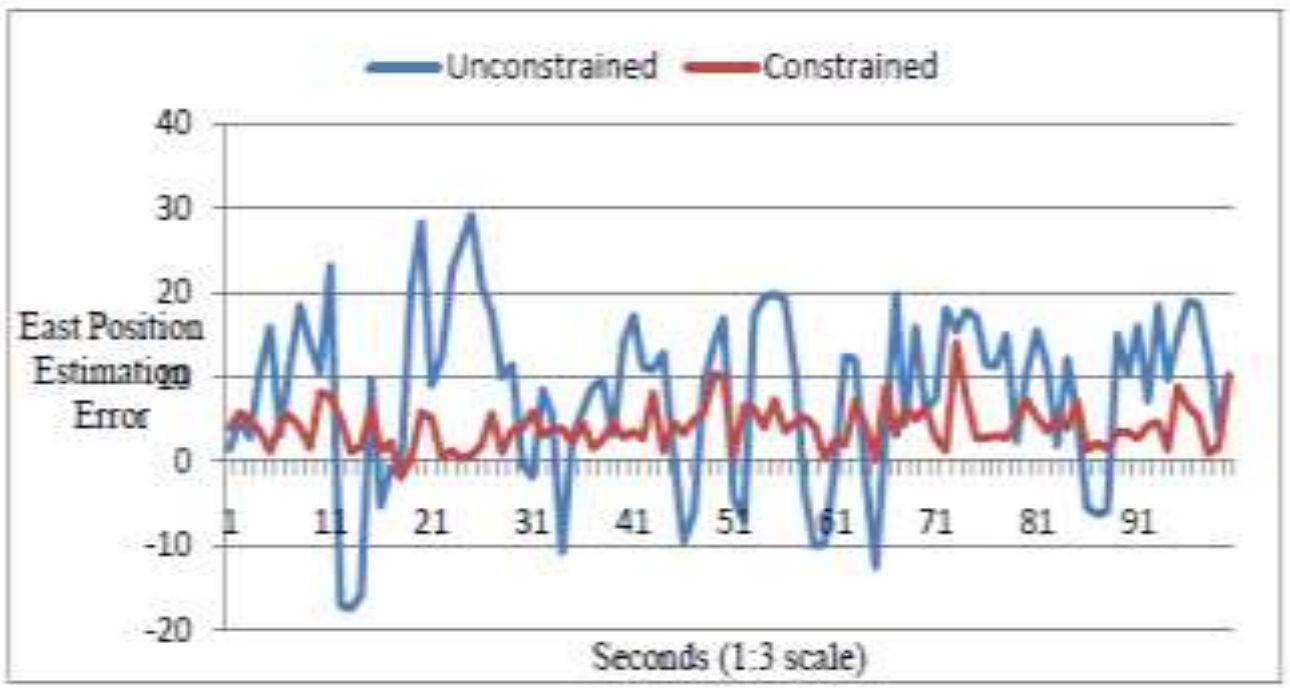

Fig. 8 Unconstrained and constrained KF in the East position estimation error.

The algorithm of position prediction and tracking with GPS is simulated with beam width $10^o$-$20^o$ and 15 seconds silent duration in 10 different experiments taking 10 Bernoulli trails at different position inside the predicted tracking zone on each experiment. The results show that the average efficiency of the proposed algorithm is approximately 99% whereas the existing algorithm has 96% as shown in Fig. 9.

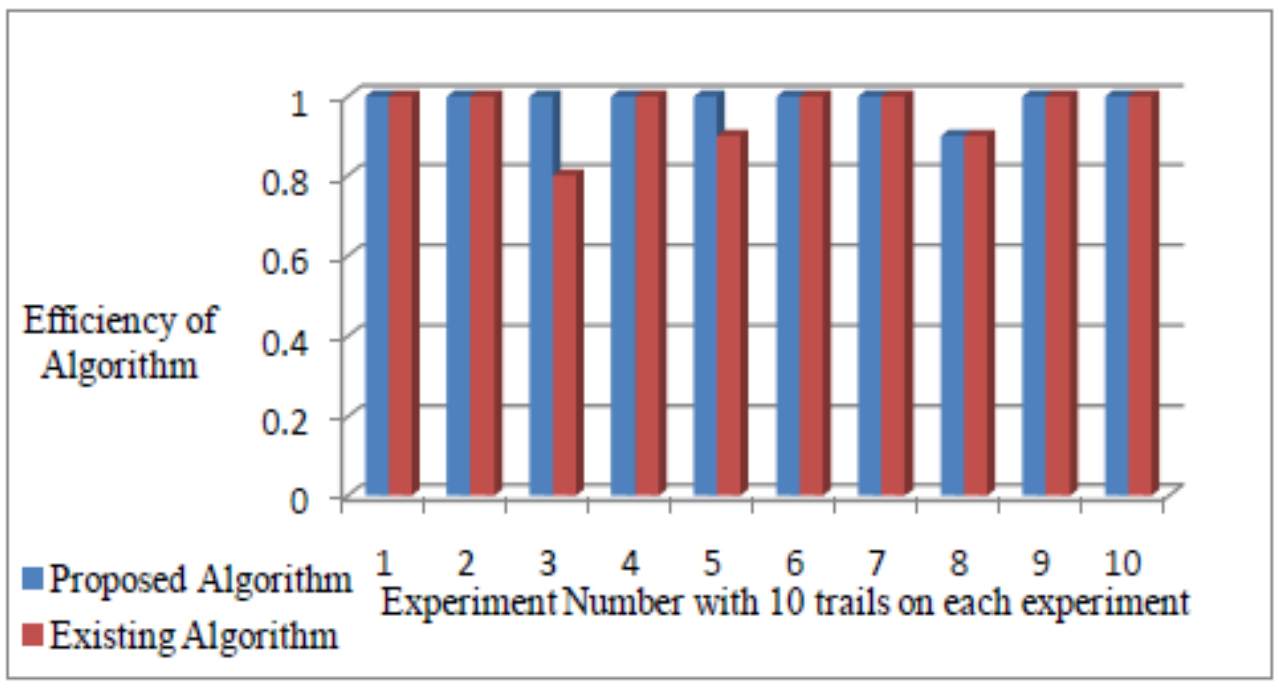

Fig. 9 Efficiency comparison of Proposed and Existing Algorithm

Similarly, while simulating the efficiency with beam width, the efficiency is found directly proportional to beam width for the battlefield characteristics of directional change about 30 degree. The efficiency of 0.7 is found at 2 degree because the coverage area is very small in such case. The efficiency is 0.95 at the beam width between 7 to 14 and 1 between 28 and 30. The proposed algorithm (Integrated Zone based PL&T) has higher efficiency against beam width as compared with the existing algorithm (LAPDC) as shown in Fig. 10.

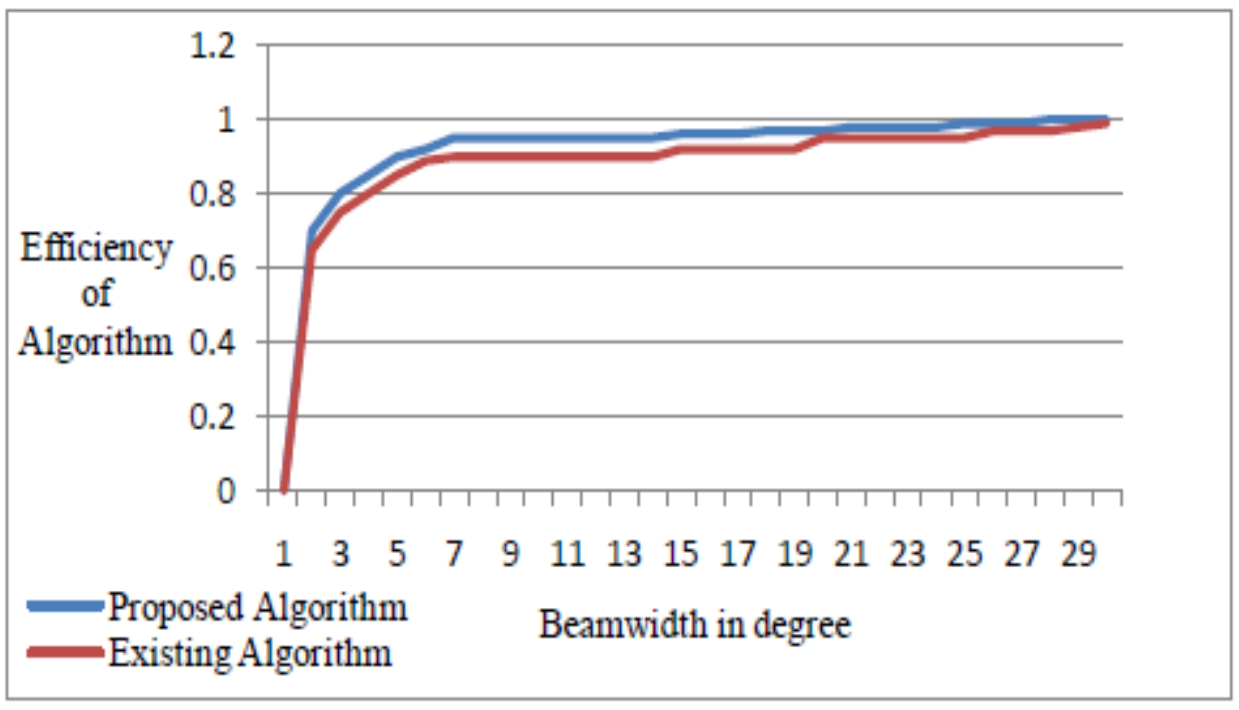

Fig. 10 Efficiency versus beam width

Additionally, the efficiency is found inversely proportional to the silent period. In battlefield a vehicle moving with average velocity of 60km/h, can covers more than 330 meters during 20 seconds or 500 meters during 30 seconds and can change its direction with an absolute relative direction angle being lager than 30 degree. The efficiency decreases to 0.38 with increasing the silent period to 30 Seconds. Again, the proposed algorithm has higher efficiency against silent period as compared with the existing algorithm as shown in Fig. 11.

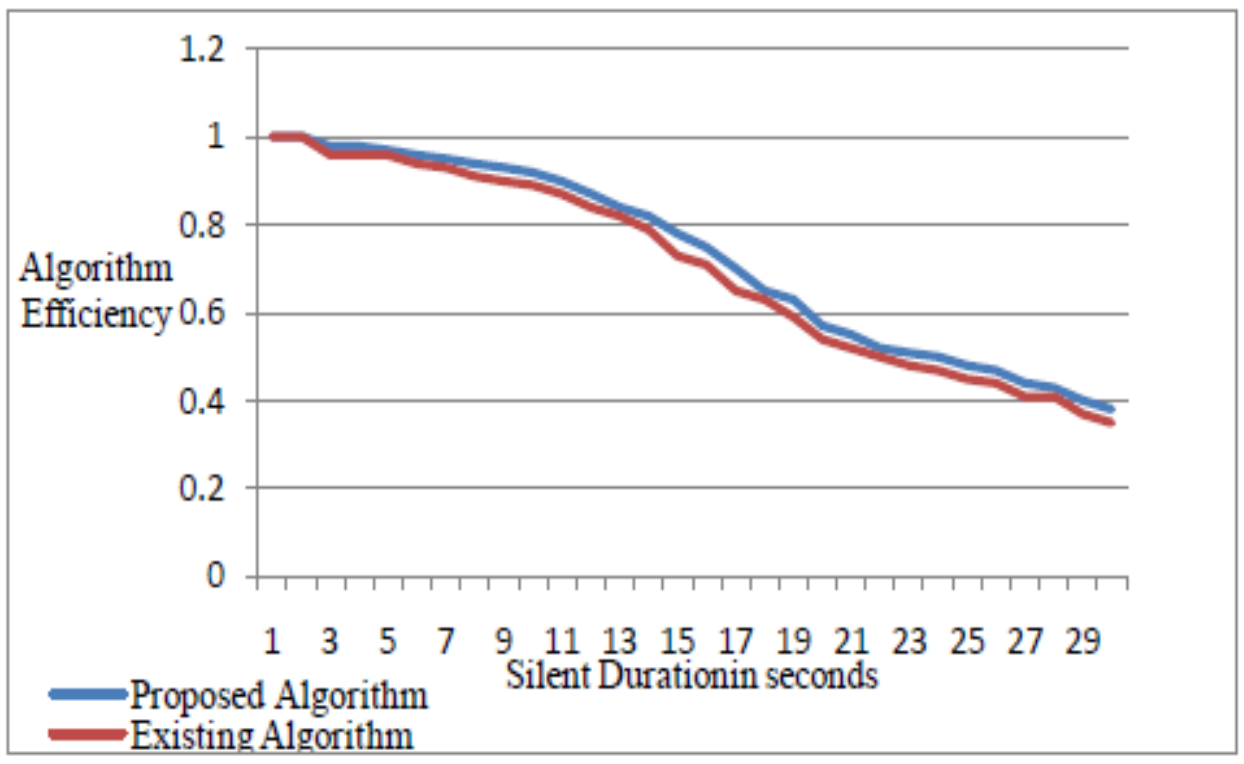

Fig. 11 Efficiency versus silent duration

### 3.1 Performance of Integrated Zone based PL&T in OFDM system using Interleaving-KV Transform Coding

Integrated Zone based PL&T is deployed in OFDM (Orthogonal Frequency Division Multiplexing) based IEEE 802.11(a) system specification as illustrated in Table-1 for robust performance in the dispersive channels. The OFDM based system is deployed at the physical layer with Integrated Zone based PL&T application. The OFDM parameters for simulation are listed in the Table-1, to analyze the Bit Error Rate (BER) performance against $E_b/N_o$ (bit energy/noise). First of all, transmitter runs the PL&T algorithm at application layer to localize and track the desired receiver sending KV blocks stream to the receiver after handshaking. KV transform coding is deployed at the physical layer with OFDMA to enable the radio channel to handle multi-path fading by recovering the data with low Bit Error Rate (BER) at low SNR or $E_b/N_o$. The transmitter side of the KV transform coding convert a set of bits to a discrete sample. Each KV transform uses four discrete samples to produce four coefficient samples at the output and two overhead samples for error correction. By modifying these real time samples, the transmission samples will have finite voltages that can be coded by discrete codes. The number of bits per transmission sample is same as the number of bits/sample at the input. The transmitter side uses multiple KV blocks each producing the transmission samples and these samples are interleaved in packets. Then, these packets are sent through OFDM based transmitter, channel (Rayleigh and Rician) and OFDM receiver. The KV transform at the decoder receives the estimated coefficient samples after sample correction and then the bits are recovered. The KV system is implemented to include the transmission of ensembles of packets, single sample error correction in each KV block, sample interleaving from a set of KV blocks, the single retransmission of selected KV blocks in error in each ensemble.

There are two different multipath channel model deployed separately known as Rayleigh channel in the absence of line of sight and Rician channel in the presence of line of sight, and then OFDM at the physical layer. The Rayleigh fading channel models that the magnitude of a signal that passed through a transmission medium will vary randomly, according to the radial component of the sum of two uncorrelated Gaussian random variables. The Rayleigh channel is model as 10 tap channel such that the real and imaginary part of each tap is an independent Gaussian random variable. On the other hand, The Rician channel has dominant line of sight in addition to Rayleigh distribution. The best and worst scenario of Rician fading channels depends upon k-factors, $k=\beta^2/2\sigma^2$ where β is the amplitude of the spectacular component σ is the variance of zero mean stationary Gaussian process. The Rician channel with K = ∞, is the Gaussian channel with strong line of sight whereas the Rician channel with K = 0, is Rayleigh channel with no line of sight path, respectively. The Rician channel with k=1 and random noise is deployed in simulation.

For multicarrier modulation, the symbol duration is 3.2 µs for subcarriers space +- 312.5KHz, +- 625KHz… and so far. The available bandwidth of 20MHz is split into 64 subcarriers. But, out of the available 64 subcarriers, only 52 subcarriers are used for transmitting the sequence of KV blocks and rest 12 subcarriers are wasted to roll up the spectrum. For this scenario, regarding the available bandwidth from -10MHz to +10MHz, only subcarriers from -8.1250MHz (-26/64*20MHz) to +8.1250MHz (+26/64*20MHz) are used. In other words, the signal energy is spread over a bandwidth of 16.25MHz but the noise is spread over bandwidth of 16.250MHz, whereas noise is spread over bandwidth of 20MHz. The cyclic prefix is guard on 16 samples as 0.8 µs from the end of the sinusoidal appended to the beginning of the sinusoidal to mitigate the natural time dispersion among symbols. This prevents signal discontinuities and achieves the original sinusoidal of frequency 312.5 KHz.

Table 1: IEEE 802.11a Specification Parameters

| **Parameter** | **Value** |
|---|---|
| Modulation | BPSK |
| FFT size | 64 |
| No. of used Subcarrier ($nD_{SC}$) | 52 |
| FFT sampling frequency | 20 MHz |
| Subcarrier spacing | 312KHz |
| Used subcarrier index | {-26 to -1, +1 to +26} |
| Cyclic prefix duration ($T_{cp}$) | 0.8µs |
| Data Symbol duration ($T_d$) | 3.2µs |
| Total Symbol duration($T_s$) | 4µs |

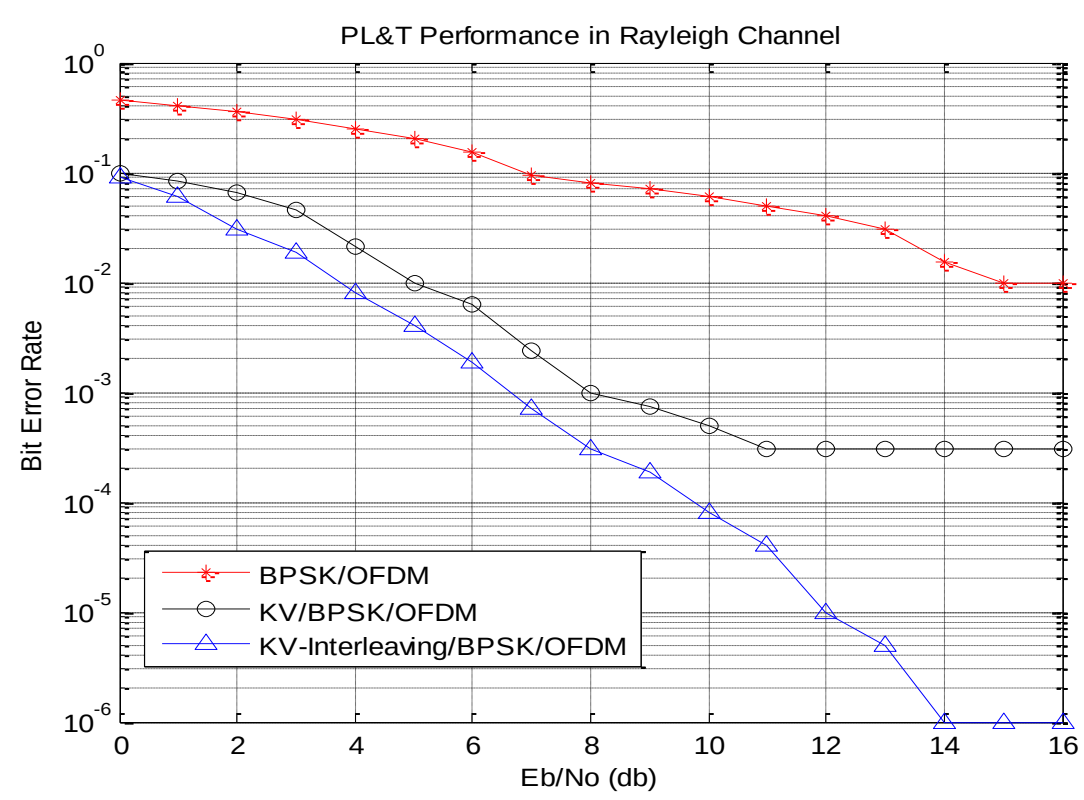

Fig. 12 PL&T performance in Rayleigh Channel

PL&T is deployed by encoding PL&T data (distance and direction) bits with three different modulation schemes categorized as BPSK/OFDM, KV/BPSK/OFDM and KV/Interleaving/BPSK/ OFDM in both Rayleigh and Rician channel. The BER is found inversely proportional to the $E_b/N_o$ or transmit power in both channels. In other words, the BER decreases as the $E_b/N_o$ increases and then optimizes at some point due to burst errors as shown in Fig. 12-13. This refers that bit stream can be recovered perfectly in higher $E_b/N_o$ until the deleterious burst errors are appeared and after that BER cannot be increased even $E_b/N_o$ increases.

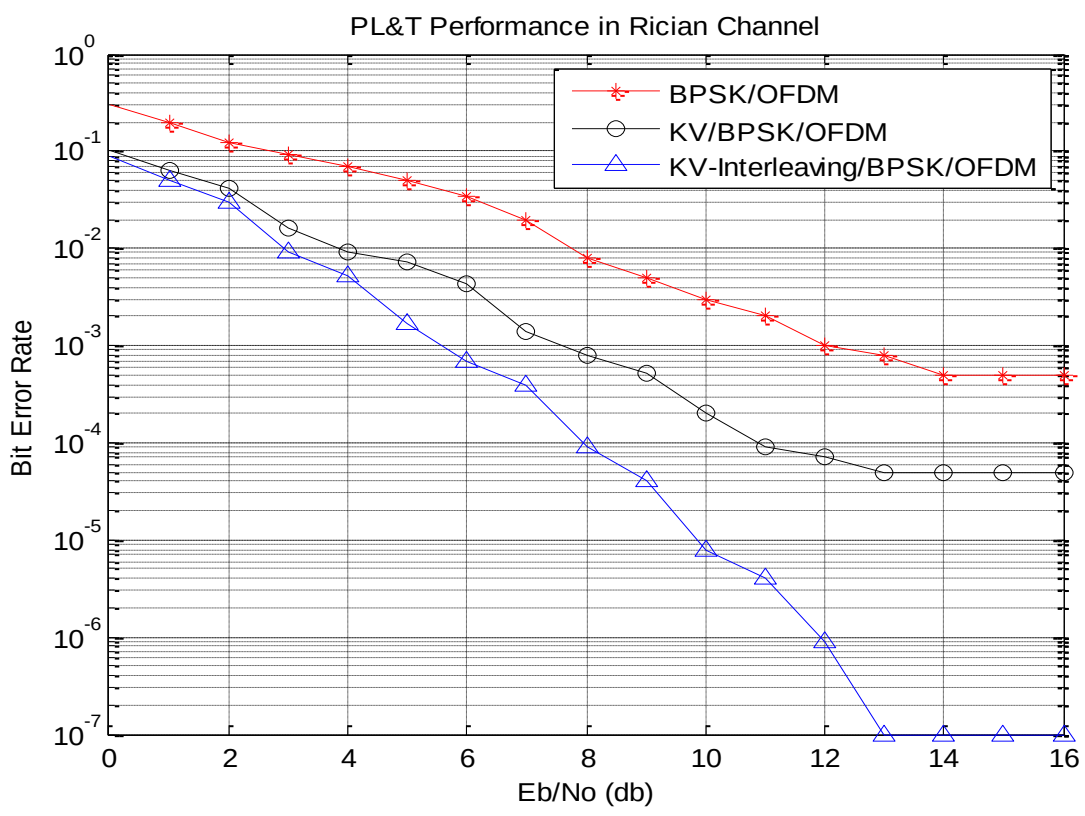

Fig. 13 PL&T performance in Rician Channel

From the simulation results, BER is found drastically reduced in KV/Interleaving/BPSK/OFDM modulation scheme in both Rayleigh and Rician channel. Therefore, KV/Interleaving/BPSK/OFDM outperforms over KV/ BPSK/OFDM and BPSK/OFDM, while KV/ BPSK/OFDM outperform over BPSK/OFDM in both Rayleigh and Rician channel as illustrated by Fig. 12-13. The major reason is that KV with interleaving in BPSK over OFDM can efficiently recover the data transmitted at receiver as it deployed the correlation between data sets during KV encoding and KV decoding. Similarly, KV in BPSK over OFDM can slightly recover data transmitted at receiver using simple error correction and data retransmission. Furthermore, the $E_b/N_o$ is found lower in Rician channel as compared to Rayleigh channel by 1 db, 2 db, 2 db,2 db, 2 db at BER of $10^{-2}$, $10^{-3}$, $10^{-4}$, $10^{-5}$, $10^{-6}$ for KV/Interleaving/BPSK/OFDM scheme. In other words, the higher BER is found in Rayleigh channel rather than Rician channel due to the higher carrier frequency offset and delay spread in the absence of line of sight which violate the orthogonality among sub carriers and increase the Inter Carrier Interference (ICI).

## 4. Conclusion

In MANET, a receiver node's real time position is predicted, estimated and tracked by a transmitter when each node has directional antenna. Real time means the location is determined inside the tracking zone before mobile node movement. In GPS free system, Baseline PL&T is deployed with extended Kalman filter considering the state constrained and unconstrained scenario for dynamic tracking. The position estimation errors are found lower in constrained filter in the North and East position estimation because there is initial state constrained for smoothing the covariance errors. Similarly, Integrated Zone based PL&T (IZPL&T) algorithm creating tracking zone, deploying the joint information of forward movement and sharp turns, has better performance than LPADC. Both the devised algorithm considers the battlefield mobility characteristics, spatial reuse, low power consumption, security and precision. The simulation illustrates the outstanding performance of IZPL&T with KV/Interleaving/BPSK/OFDM modulation scheme for both Rician and Rayleigh channel. Future research will concentrate on more secured multiple targets' PL&T with effective zonal computation in mobile ad hoc network fabrics.

**ACKNOWLEDGMENT**

This research work is supported in part by the U.S. ARO under Cooperative Agreement W911NF-04-2-0054 and the National Science Foundation NSF 0931679. The views and conclusions contained in this document are those of the authors and should not be interpreted as representing the official policies, either expressed or implied, of the Army Research Office or the National Science Foundation or the U. S. Government.

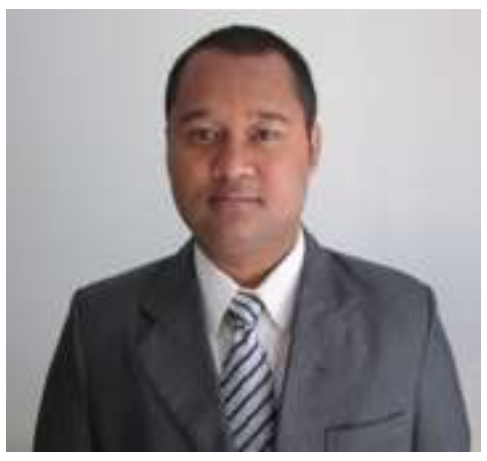

**Niraj Shakhakarmi** worked as a Doctoral Researcher since 2009-2011 in the ARO Center for Digital Battlefield Communications (CeBCom) Research, Department of Electrical and Computer Engineering, Prairie View A&M University. He received his B.E. degree in Computer Engineering in 2005 and M.Sc. in Information and Communications Engineering in 2007. His research interests are in the areas of Wireless Communications and Networks Security, Secured Position Location & Tracking (PL&T) of Malicious Radios, Cognitive Radio Networks, WCDMA/HSPA/LTE/WRAN, Next Generations Wireless Networks, Satellite Networks and Digital Signal Processing, Wavelets Applications and Image/Colour Technology. He is a member of IEEE Communications Society, ISOC, IAENG and attended AMIE conference. He has published several WSEAS and IJCSI journals, WTS, Elsevier and ICSST conference paper. His several journals and conference papers are under review in IEEE journals. He is serving as reviewer for WSEAS, SAE, WTS and editorial board member for IJEECE, WASET, JCS and IJCN journals.

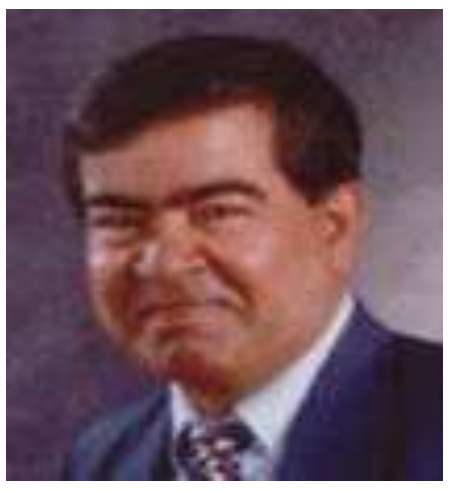

**Dhadesugoor R. Vaman** is Texas A & M University Board of Regents and Texas Instrument Endowed Chair Professor and Founding Director of ARO Center for Battlefield Communications (CeBCom) Research, ECE Department, Prairie View A&M University (PVAMU). He has more than 38 years of research experience in telecommunications and networking area. Currently, he has been working on the control based mobile ad hoc and sensor networks with emphasis on achieving bandwidth efficiency using KV transform coding; integrated power control, scheduling and routing in cluster based network architecture; QoS assurance for multi-service applications; and efficient network management.

Prior to joining PVAMU, Dr. Vaman was the CEO of Megaxess (now restructured as MXC) which developed a business ISP product to offer differentiated QoS assured multi-services with dynamic bandwidth management and successfully deployed in several ISPs. Prior to being a CEO, Dr. Vaman was a Professor of EECS and founding Director of Advanced Telecommunications Institute, Stevens Institute of Technology (1984 1998); Member, Technology Staff in COMSAT (Currently Lockheed Martin) Laboratories (1981-84) and Network Analysis Corporation (CONTEL) (1979-81); Research Associate in Communications Laboratory, The City College of New York (1974-79); and Systems Engineer in Space Applications Center (Indian Space Research Organization) (1971-1974). He was also the Chairman of IEEE 802.9 ISLAN Standards Committee and made numerous technical contributions and produced 4 standards. Dr. Vaman has published over 200 papers in journals and conferences; widely lectured nationally and internationally; has been a key note speaker in many IEEE and other conferences, and industry forums. He has received numerous awards and patents, and many of his innovations have been successfully transferred to industry for developing commercial products.